# Two-phonon contributions to inelastic x-ray scattering spectra of MgB$_2$

A. Q. R. Baron[1,2*], H. Uchiyama[3], R. Heid[4], K.P. Bohnen[4], Y. Tanaka[1], S. Tsutsui[2], D. Ishikawa[1], S. Lee[5], and S. Tajima[5,6]

[1]SPring-8/RIKEN, 1-1-1 Kouto, Sayo, Hyogo 679-5148, Japan
[2]SPring-8/JASRI, 1-1-1 Kouto, Sayo, Hyogo 679-5198, Japan
[3]Dept. of Mat. Chem., Ryukoku University, 520-2194, Japan
[4]Forschungszentrum Karlsruhe, Institut für Festkörperphysik, Box 3640, D-76021, Germany
[5]Superconductivity Research Laboratory, ISTEC, Tokyo, 135-0062, Japan
[6]Dept. of Physics, Osaka University, Osaka, 560-0043, Japan

**Abstract**

Two-phonon contributions to meV-resolved inelastic x-ray scattering spectra of MgB$_2$ at 300K are identified, in good agreement, in both intensity and energy, with a harmonic calculation using the force constant matrix from *ab-inito* LDA calculations. This contribution impacts the determination of the linewidth of the E$_{2g}$ phonon mode that is so important for the high T$_c$ of this material. To the best of our knowledge, this is the first observation of peaks in measurements of phonon dispersion (q>0) due to 2-phonon scattering in a non-rare-gas solid.

[*] baron@spring8.or.jp

MgB$_2$ with its high T$_c$ [1] of nearly 40K has generated an immense amount of interest over the past 5 years. It is now generally accepted that MgB$_2$ is a phonon-mediated BCS/Eliashberg superconductor with extremely strong electron-phonon coupling to the phonon modes with E$_{2g}$ symmetry at Γ. As such, the phonon spectra and dispersion have generated significant attention using a variety of methods. However, there remain some subtleties of the phonon spectra that have not yet been entirely understood. One is the difference between Raman measurements [2-4] with expectations from calculation [5], another is the appearance of anomalous phonon modes in inelastic x-ray scattering (IXS) spectra [6].

Here we show that the 2-phonon contribution to IXS spectra of MgB$_2$ is significant, generating peaks in the data that mimic the effect of 1-phonon modes, overlapping interesting modes, and affecting the evaluation of the linewidth of the crucial E$_{2g}$ mode. While, previously, a similar effect was observed in inelastic neutron scattering from rare-gas solids [7], it is rather surprising to see peaks from 2-phonon contributions in a more conventional material, well below its melting point. However, the use of a relatively new and nearly background-free technique, IXS, facilitates this observation.

IXS differs from inelastic neutron scattering (INS), the conventional method for investigation phonon dispersion, in that there are fewer sources of background. In neutron scattering, inelastic backgrounds appear from incoherent scattering, multiple scattering, and multi-phonon contributions. These usually lead to the subtraction of a slowly varying function of energy from INS spectra. In contrast, IXS measurements do not have contributions from incoherent scattering or multiple scattering: for x-rays, incoherent, Compton, scattering occurs with larger (>eV) energy transfer due to the small electron mass, while multiple scattering is precluded by the high photoelectric absorption cross-section of most materials. The multi-phonon part is then typically the only intrinsic background in IXS for ~meV energy transfers. While one might hope, based on INS data treatment, that the multi-phonon contribution would be some smooth function, easily subtracted out, in fact calculations quickly show that it can have noticeable structure, and thus, in principle, can be a significant confound in carefully analyzed experiments with high-quality data.

In previous work [6], contributions to IXS spectra of MgB$_2$ were observed in a pure longitudinal geometry that appeared to be anomalous since phonons at the relevant energies were forbidden from contributing to the spectra by symmetry (note also [8]). As the dispersion was mostly flat, it was speculated [6] the intensity was from localized modes due to a Mg-deficiency, as was suggested to be present by structural measurements [10]. We now understand that the strong additional peak in these spectra is due to multi-phonon contributions, as it agrees nicely with 2-phonon calculations using a harmonic expansion of eigenmodes derived from the force constant matrix from ab-initio simulation, and is appropriately temperature dependent.

Within a harmonic model, the dynamic structure factor, S(**Q**,ω) can be expanded in a power series giving terms corresponding to Bragg (no-phonon) scattering, 1-phonon scattering, 2-phonon scattering, etc. (see [11,12] for detailed expositions, or [13] for an abbreviated treatment focusing on x-ray scattering. More generally, anharmonic processes can lead to interference effects [14]) The 1-phonon contribution for a crystalline material with r atoms/unit cell can be written [11],

$$S(\mathbf{Q},\omega)_{1p} = N \sum_{\substack{\mathbf{q} \\ 1st\ Zone}} \sum_{\substack{j \\ 3rModes}} \frac{1}{\omega_{\mathbf{q}j}} \{S_+ + S_-\}$$

$$S_+ \equiv \left| \sum_{\substack{\mathbf{d} \\ Atoms/Cell}} f_d(\mathbf{Q})\ e^{-W_d(\mathbf{Q})}\ \frac{\mathbf{Q}\cdot\mathbf{e}_{\mathbf{q}jd}}{\sqrt{2M_d}}\ e^{-i\mathbf{Q}\cdot\mathbf{d}} \right|^2 \langle n_{\mathbf{q}j}+1 \rangle\ \delta_{\mathbf{Q}-\mathbf{q},\tau}\ \delta(\omega-\omega_{\mathbf{q}j})$$

(1)

**Q** is the total momentum transfer, **q** is a momentum in the first Brillouin zone, $\hbar\omega$ is the energy transfer to the lattice, and j is the index (1 to 3r) of the mode with energy $\hbar\omega_{\mathbf{q}j}$ and polarization $\mathbf{e}_{\mathbf{q}jd}$. N is the number of unit cells, $f_d(Q)$, $M_d$, $e^{-2W_d(\mathbf{Q})}$ are the, x-ray atomic form factor, mass, and Debye-Waller factor for the $d^{th}$ atom in the unit cell (located at **d**), and $\tau$ is the nearest reciprocal lattice vector. $S_+$ and $S_-$ are terms describing phonon creation and annihilation, respectively. $S_-$ is obtained from $S_+$ by changing the sign in front of **q** in the Kronecker delta and $\omega_{\mathbf{q}j}$ in the Dirac delta, dropping the "+1" in the occupation term and taking the complex conjugate of the phonon polarization, $\mathbf{e}_{\mathbf{q}jd}$.

The 2-phonon contribution may be written (based on [11]) as a sum of four terms,

$$S(\mathbf{Q},\omega)_{2p} = \frac{1}{2} N\ \frac{\hbar}{N_q} \sum_{\substack{\mathbf{qq'} \\ 1st\ Zone}} \sum_{\substack{jj' \\ 3rModes}} \frac{1}{\omega_{\mathbf{q}j}} \frac{1}{\omega_{\mathbf{q'}j'}} \{S_{++} + S_{+-} + S_{-+} + S_{--}\}$$

$$S_{++} = \left| \sum_{\substack{\mathbf{d} \\ Atoms/Cell}} f_d(\mathbf{Q})\ e^{-i\mathbf{Q}\cdot\mathbf{d}} e^{-W_d}\ \frac{\mathbf{Q}\cdot\mathbf{e}_{\mathbf{q}jd}\ \mathbf{Q}\cdot\mathbf{e}_{\mathbf{q'}j'd}}{2M_d} \right|^2$$

$$\times\ \langle n_{\mathbf{q}j}+1 \rangle \langle n_{\mathbf{q'}j'}+1 \rangle\ \delta_{\mathbf{Q}-\mathbf{q}-\mathbf{q'},\tau}\ \delta(\omega-\omega_{\mathbf{q}j}-\omega_{\mathbf{q'}j'})$$

(2),

corresponding to 2-phonon creation, $S_{++}$, creation and annihilation, $S_{+-}$ and $S_{-+}$, and 2-phonon annihilation, $S_{--}$. $N_q$ is the number of terms used in the sum over **q**. $S_{++}$ is given explicitly, with other terms ($S_{+-}$, etc) obtained from the transformation given above, with the first subscript referring to the unprimed quantities, **q**, j , and the second referring to the primed ones, **q'**, j'.

The 2-phonon contribution may be calculated using the force constant matrix from *ab-initio* calculations. In this case, we have used a matrix extending out to 33 Å bond lengths based on interpolation from an LDA mixed-basis pseudo-potential calculation similar to that in [15][6]. Details about the pseudopotentials and local functions can be found in [16]. In order to get accurate results, dynamical matrices were obtained on a hexagonal (18x18x6) q-point mesh.

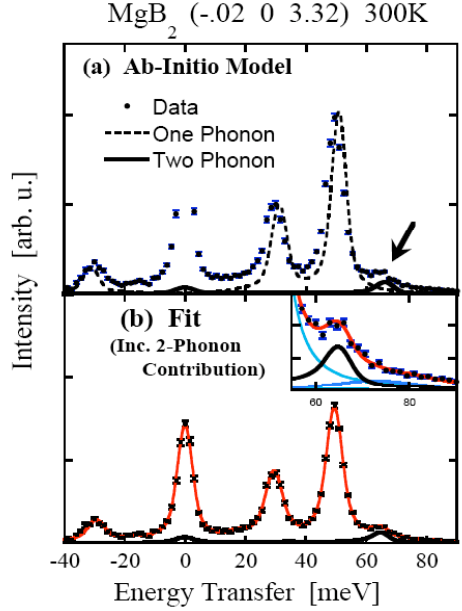

**Fig. 1.** Comparison of data and calculation. In (a) the ab-initio results for the one and 2-phonon contributions are shown, while (b) shows a fit with the 1-phonon energies allowed to shift, but the 2-phonon contribution held in fixed ratio to the 1-phonon one. See text.

Figure 1 compares data with calculations, demonstrating the presence of the 2-phonon contribution. The data was measured at BL35XU [18] of SPring-8 using 6 meV resolution at 15.816 keV (Si (888) reflection). The sample was a small c-axis normal platelet of $MgB_2$ about 0.2 x 0.4 x 0.05 $mm^3$ grown as described in [19], and held in vacuum at room temperature. The geometry, nearly purely longitudinal along ΓA, has been chosen so that only two strong phonons are present. The solid lines are calculations using eqns. 1 and 2 with the phonon mode energies and eigenvectors determined from the force-constant matrix of the ab-initio LDA calculations and convolved with the resolution of the spectrometer. It is worth emphasizing that, while there is an over-all scale factor that has been chosen to get good agreement between the calculation and the data, the relative intensity of the one and 2-phonon contributions is not free, being fixed by eqns 1 and 2. One can immediately see that the peak in the data at 67 meV, which is not present in the 1-phonon calculation, is well described in both intensity and position by the 2-phonon calculation. Figure 1b then shows a fit to the data, including the 1-phonon lines expected from the *ab-initio* calculation (but allowed to shift to match the data) and the 2-phonon contribution directly seen in 2b, and some elastic background. The intensity of the 2-phonon term has been fixed to a constant fraction of the 1-phonon intensity as given by the calculation used in Fig. 1a. The agreement is seen to be excellent [20].

The strong 2-phonon contribution at ~67 meV is due to simultaneous excitation of two acoustic phonon modes (the $S_{++}$ term in eqn. 2). In Fig. 2 we plot the evolution of the 2-phonon component in two symmetry directions, along ΓA and along ΓM. In both cases, the predominant intensity is in the region from 60 to 75 meV energy transfer, due to simultaneous excitation of two acoustic modes. This is not surprising as acoustic eigenvectors generally follow the simplest phase relationship and allow the largest dot-product with the momentum transfer. Optic modes involve out-of-phase motions of atoms in the cell, and, by virtue of cancellation of terms in the sum over the unit cell (and their higher frequency), tend to be weaker. The fact that the intensity is peaked in 60 to 75 meV range just comes from the q-integration in eqn 2: there is a lot of the zone where the acoustic modes have flattened out at about 30 to 40 meV, so the contribution at double this value is correspondingly large – somewhat analogous to a van-Hove singularity in the density of states. The peak near zero energy transfer is predominantly due to simultaneous creation and annihilation of acoustic modes ($S_{+-}$ and $S_{-+}$ in eqn. 2). We note that the structure visible in figure 2 is in fact very different than the 2-phonon density of states sometimes discussed in Raman scattering (derived from eqn. 2 by ignoring the frequency denominators and taking the sum over atom sites to unity and setting Q=0). Proper consideration of the phonon polarizations, as is done here, strongly increases the relative contribution of the peak at ~67 meV.

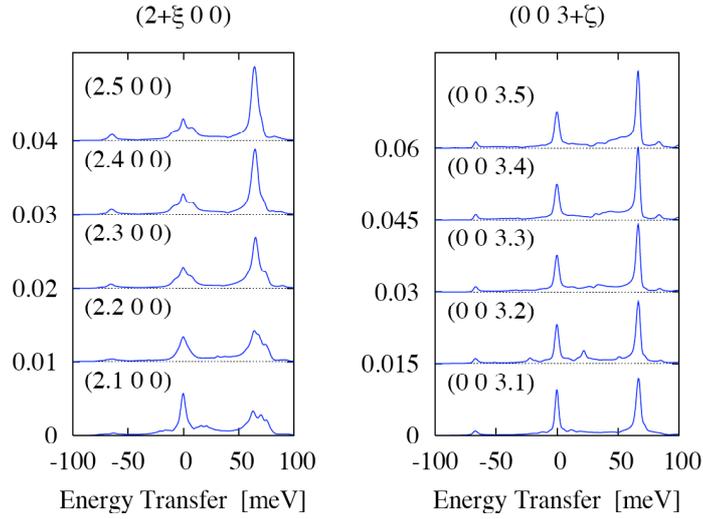

**Fig. 2.** Calculated 2-phonon contribution at 300K in different symmetry directions.

Further confirmation of the nature of the ~67 meV contribution is obtained from its temperature dependence. 1-phonon intensities should vary (eqn. 1) as the usual bose factor $\langle 1+n_j \rangle = \left(1 - e^{-\hbar\omega_j/k_B T}\right)^{-1}$, while the 2-phonon term is more complicated (eqn. 2), and is generally more temperature sensitive. We measured spectra near (0 0 3.5) at 300 and 16K, and found that the integrated intensity in the 60 to 70 meV region changed strongly from 0.19±0.02 at 300K to 0.11±0.01 at 16K (normalized to a clear and strong 1-phonon line at 48 meV), a reduction of about 42%. If we were to assume the intensity in the 60 to 70 meV region is due to processes obeying 1-phonon statistics, we would expect an *increase* in relative intensity of about 10%, due primarily to reduction of the intensity of the 48 meV mode used for normalization. Meanwhile calculations using eqn. 1 and 2, suggest the two-phonon contribution would drop from 0.21 (300K) to 0.13 (16K), a 38% decrease, in reasonable agreement with the measurement. This confirms our interpretation that the intensity in this region comes from two-phonon contributions.

We now consider the spectra along Γ-M in a purely longitudinal geometry. This is the region where one can expect the $E_{2g}$ mode to contribute: as one moves from the outer part of the Brillouin zone, M, toward Γ, within the plane, the $E_{2g}$ mode both strongly softens, in analogy with a Kohn anomaly, and broadens, approximately in agreement with calculations [6]. Figure (3a) shows the results of direct calculation based on the *ab-inito* model, including convolution of a Lorentzian to approximate the experimental resolution. It is clear that modeling the spectrum by just 1-phonon modes would lead to a fit that would tend to over-estimate the width of the $E_{2g}$ mode, especially as the fit would allow the amplitude of the low-energy (55 meV) mode to be free. Fits to the data are shown in figure 3b. The procedure was the same as for Fig. (1b) and good fits were obtained, with the new linewidth parameters plotted in Fig. 4 along with the calculation and the results from [6]. In particular, the experimental estimate at 0.45Å$^{-1}$ is now in much better agreement with calculation. We also note a general trend toward a measured linewidth larger than calculation at small momentum transfers. Here we do not comment further, but suggest additional data with improved energy resolution is needed to determine the linewidth accurately. The 2-phonon contribution might also partially account for the difference between the results of [6] and [9] when analyzing data along Γ-A in a transverse geometry (see discussion in [6]).

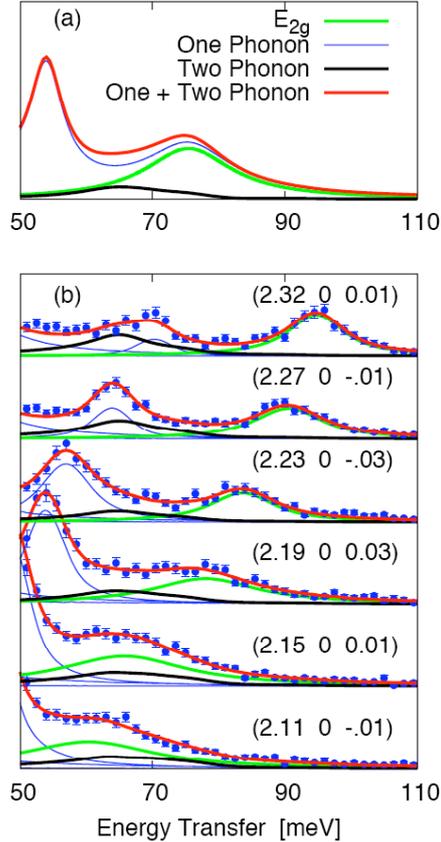

**Fig. 3.** Spectra along Γ-M. Note the 2-phonon contribution (black) is not negligible as compared to the $E_{2g}$ mode (green). Pure calculation is shown in (a) at (2.19 0 0) while data with fits are shown in (b). See text for details.

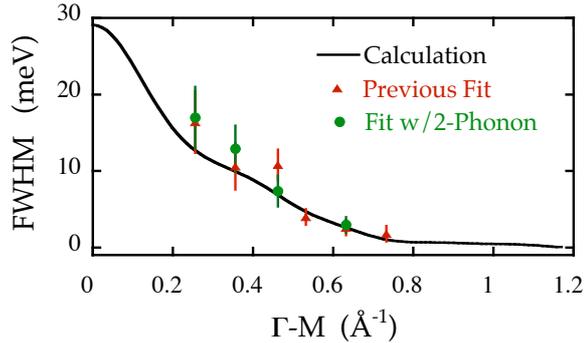

**Fig. 4.** Calculated and Measured $E_{2g}$ linewidth. Black line and red points are from [6] while the green points are the present results. See text.

Finally we turn to the disagreement between calculation and measurement for the Raman data mentioned in the introduction. As pointed out in [5] the momentum transferred to the $E_{2g}$ mode in Raman work is not sufficient to move holes far enough from the Fermi-surface to absorb the full energy (>60 meV) of the mode by an intraband electronic excitation. While finite temperature allows some freedom, at the level of $k_BT$, numerically this is found still to be insufficient for the relatively high energy of the $E_{2g}$ mode [5]. This reduction of Landau damping at low Q has been well discussed in the literature (see, e.g., [19]), and would be expected to lead to a drastic narrowing of the $E_{2g}$ linewidth in Raman scattering relative to IXS measurements and calculations at larger Q [5]. In contrast, what is observed [2-4] is a linewidth broadening at Γ. While some recent work has suggested this could be caused by the Raman scattering probing the phonon lifetime *and* the relaxation of screening carriers [22], here, in light of our IXS data, we suggest that a 2-phonon contribution might also be important. The relevant cross section involves a different calculation than that described here (see [23][24]), but one might expect that the 2-phonon contribution could be non-negligible, compared to the $E_{2g}$ mode. This contribution might also play a role in explaining why the strong temperature dependence predicted in [25,22] is not observed in the IXS spectra [6], though other effects may also matter [26].

In sum, we have shown that 2-phonon contributions are significant in the IXS spectra of $MgB_2$. The nearly background-free nature of the IXS technique allows one to clearly see this contribution, and we confirm the intensity and energy of the contribution is in good agreement with harmonic calculations. The temperature dependence of the contribution does not agree with the single phonon bose factor, but is in reasonable agreement with 2-phonon estimates. The contribution is seen to affect the determination of the $E_{2g}$ linewidth, bringing the measurements in closer agreement with calculations at one momentum transfer.

This work was carried out at SPring-8 under proposals No. 2002A0559 and 2002B0594 . This work was partially supported (through ISTEC-SRL) by the New Energy and Industrial Technology Department Organization (NEDO) as Collaborative Research and Development of Fundamental Technologies for Superconductivity Applications.